\documentclass[journal=jascat,manuscript=article]{achemso}

\usepackage{chemformula} 
\usepackage[T1]{fontenc} 

\usepackage[numbers,sort&compress]{natbib}
\usepackage{soul}
\usepackage{soulpos}
\sethlcolor{yellow}
\usepackage[markup=underlined]{changes}
\definecolor{delcolor}{RGB}{180,0,0}
\setdeletedmarkup{\textcolor{delcolor}{\sout{#1}}}

\author{Amirhossein Fallah}
\affiliation{Department of Electrical and Computer Engineering, University of Southern California, Los Angeles, California 90089, United States}
\author{Constantine Sideris}
\affiliation{Department of Electrical Engineering, Stanford University, Stanford, California 94305, United States}
\email{sideris@stanford.edu}

\title[JVIE Inverse Design]
  {Fast 3D Nanophotonic Inverse Design using Volume Integral Equations}


\begin{document}


\begin{abstract}
Designing nanophotonic devices with minimal human intervention has gained substantial attention due to the complexity and precision required in modern optical technologies. While inverse design techniques typically rely on conventional electromagnetic solvers as forward models within optimization routines, the substantial electrical size and subwavelength characteristics of nanophotonic structures necessitate significantly accelerated simulation methods. In this work, we introduce a forward modeling approach based on the volume integral equation (VIE) formulation as an efficient alternative to traditional finite-difference (FD)-based methods. We derive the adjoint method tailored specifically for the VIE framework to efficiently compute optimization gradients and present a novel unidirectional mode excitation strategy compatible with VIE solvers. Comparative benchmarks demonstrate that our VIE-based approach provides multiple orders of magnitude improvement in computational efficiency over conventional FD methods in both time and frequency domains. To validate the practical utility of our approach, we successfully designed three representative nanophotonic components: a 3 dB power splitter, a dual-wavelength Bragg grating, and a selective mode reflector. Our results underscore the significant runtime advantages offered by the VIE-based framework, highlighting its promising role in accelerating inverse design workflows for next-generation nanophotonic devices.
\end{abstract}

\section{Introduction}
Silicon photonics has attracted significant attention as a rapidly growing research field in recent years, due to its expanding range of applications, including optical biosensors\cite{Biosensor, Biosensor2D}, quantum computing\cite{Quantum}, high-speed optical interconnects\cite{Interconnect}, LiDAR (Light Detection and Ranging)\cite{Lidar}, and optical neural networks\cite{OpticalNN_Dirk, OpticalNN}. A crucial aspect of this technology involves the design of high-performance nanophotonic component building blocks, whose performance is critical for meeting the required specifications for each application. The lack of analytical solutions and design intuition for these complicated structures precludes first principle design approaches that rely heavily on human intuition. As a result, inverse design methods have been developed and used very successfully in recent years to automate the design of such devices~\cite{Inverse_Design, Inverse_Grating, Vuc_Demul, Vuc_Const, WGF_2D, Grating_Coupler, WGF_3D}. Inverse design aims to improve the performance of nanophotonic structures using optimization algorithms\cite{Adam, PSO, Nesterov, LBFGS}, which require a forward solver to simulate the device at each iteration. As such, the forward solver is almost always the rate-limiting step that determines the overall speed of the inverse design process, and it is therefore very important to use fast, accurate, and robust solvers. In the photonic frequency regime, full-wave electromagnetic simulation is particularly challenging due to the large electrical size of devices and the presence of subwavelength features.

Finite-Difference-based methods have been widely used in simulating nanophotonic structures, both in the time domain (FDTD) and frequency domain (FDFD). Although FD methods can be quite versatile, they also suffer from significant numerical dispersion, which accumulates and can be exacerbated for long structures that span many wavelengths. FDFD methods produce sparse linear systems of equations that must be solved either using iterative or sparse direct matrix solvers. Due to their poor conditioning, prohibiting effective use of iterative solvers and the excessive memory requirements of direct solvers, FDFD methods are typically unsuitable for solving problems spanning more than a few wavelengths in size. On the other hand, although FDTD, which is a time-domain method, can be used to solve significantly larger problems than FDFD, it suffers from increased discretization errors due to the finite-difference approximation of both the temporal and spatial derivatives.  Moreover, the duration of the simulation depends on the number of time steps required for the energy inside the domain to dissipate below a threshold tolerance, which grows as the size of the simulation domain increases.

Integral equation methods provide an alternative approach for full-wave electromagnetic simulation in the frequency domain. When considering the topology optimization and shape optimization of large-scale 3D nanophotonic structures, volume integral equation (VIE) methods, in particular, are particularly attractive. A number of different VIE formulations and discretization approaches have been proposed in the literature\cite{VIE, VIE_higher, VIE_Aniso, VIE_Markanen}. Unlike other VIE formulations that require divergence-confirming or curl-confirming basis functions~\cite{VIE_Markanen} due to the unknowns representing the electric or magnetic fields, the electric current density volume integral equation \cite{JVIE} (JVIE) imposes no restrictions on the choice of basis functions for discretization. Furthermore, JVIE is a second-kind Fredholm integral equation that remains well-conditioned for high-frequency scenarios and problems with moderate index contrast. For the large problem sizes considered in nanophotonics, the discrete JVIE matrix system must be solved using iterative methods \cite{QMR, CG, GMRES}, in which the most computationally expensive operation is the matrix-vector product (MVP) of the JVIE system matrix against a vector due to the dense matrix nature of integral equation methods. For a system with $n$ unknowns, a simple MVP requires $O(n^2)$ operations. However, a key advantage of the JVIE formulation lies in the Toeplitz-like structure of its system matrix~\cite{JVIE}, which allows the MVPs to be accelerated to $O(n\log n)$ by leveraging Fast Fourier Transforms (FFT), leading to significant speed improvement, particularly for large systems. Moreover, it has been shown that using appropriately constructed preconditioners, JVIE solvers can achieve rapid convergence even for structures that are electrically large in one or two dimensions \cite{Circ_Prec}, which are common in practical photonic designs. As a result of these advantages, the JVIE method becomes a highly promising solver for large-scale 3D nanophotonic simulations.

In this work, we demonstrate the inverse design of nanophotonic structures using the JVIE as a forward model. To enable this, we introduce and incorporate mode sources, mode monitors, and gradient calculation using the adjoint method within the context of the JVIE environment. We demonstrate the advantages of using the JVIE in the inverse design process, particularly for large-scale 3D simulations.
In section 2, we present the mathematical formulation of the JVIE method and describe the implementation of mode sources to excite the desired mode in each simulation. We also compare the performance of JVIE against commonly used finite-difference solvers. In section 3, the setup and inverse design process of two nanophotonic structures, a power splitter and a dual-wavelength Bragg grating, are demonstrated. The full mathematical details of the adjoint gradient computation and mode source construction, as well as an additional inverse design example of a selective mode reflector are provided in the Supporting Information.

\section{Volume Integral Equation Formulation and Comparison with Finite-Difference Methods}

In this section, modeling three-dimensional nanophotonic structures using the JVIE formulation is introduced, and JVIE is compared against finite-difference methods to highlight several of its advantages.

\subsection{JVIE Solver}
Considering a non-magnetic, dielectric object excited by an incident electromagnetic wave, the equivalent electric current density in terms of the total electric field $\mathbf{E}$ can be defined as $\mathbf{J}_{\text{eq}}(\mathbf{r}) = j \omega \varepsilon_0 (\varepsilon_r(\mathbf{r}) - \varepsilon_r^{BG}) \mathbf{E}(\mathbf{r})$, where $\omega$ is the angular frequency, $\varepsilon_0$ is the free-space permittivity, $\varepsilon_r(\mathbf{r})$ is the relative permittivity of the dielectric, and $\varepsilon_r^{BG}$ is the relative permittivity of the background medium. Starting from Maxwell's equations and subtracting the incident fields $\mathbf{E}_{\text{inc}}(\mathbf{r})$ generated by external sources, the electric fields $\mathbf{E_{\text{s}}}(\mathbf{r})$ scattered from the dielectric object satisfy:
\begin{equation}
\nabla\times\nabla\times\mathbf{E_{\text{s}}}(\mathbf{r}) - {k_{\text{BG}}}^2\mathbf{E_{\text{s}}}(\mathbf{r}) = -j\omega\mu_0\mathbf{J_{\text{eq}}}(\mathbf{r})
\end{equation}

Here $k_{BG}$ is the wavenumber of the background medium. Then, using the dyadic Green's function $\bar{\mathbf{\Gamma}}(\mathbf{r},\mathbf{r'}) = (\bar{\mathbf{I}} + \frac{1}{{k_{\text{BG}}}^2}\nabla\nabla)
\frac{e^{-jk_{\text{BG}}|\mathbf{r} - \mathbf{r'}|}}{4\pi|\mathbf{r}-\mathbf{r'}|}$, the electric field volume integral equation (EFIE) is obtained:
\begin{equation}
\mathbf{E}(\mathbf{r}) = \mathbf{E_{\text{inc}}}(\mathbf{r}) + {k_{\text{BG}}}^2 \int{\bar{\mathbf{\Gamma}}(\mathbf{r},\mathbf{r'}).(\varepsilon_r(\mathbf{r}) - \varepsilon_r^{BG})\mathbf{E}(\mathbf{r'}) d\mathbf{r'}}
\end{equation}
Finally, using the definition of $\mathbf{J}_{\text{eq}}$, it can been shown that the following equation is satisfied\cite{JVIE,Volakis}:
\begin{equation}
(1 - \mathcal{M} \mathcal{N}) \mathbf{J}_{\text{eq}}(\mathbf{r}) = j \omega \varepsilon_0 \mathcal{M} \mathbf{E}_{\text{inc}}(\mathbf{r})
\end{equation}
Here, the operators $\mathcal{M}$ and $\mathcal{N}$ are defined as $
\mathcal{M}(\mathbf{J}(\mathbf{r})) = \frac {\varepsilon_r(\mathbf{r}) - \varepsilon_r^{BG}}{\varepsilon_r(\mathbf{r})}
\mathbf{J(\mathbf{r})}$,
$
\mathcal{N}(\mathbf{J}) = \nabla \times \nabla \times\int \frac{e^{-jk_{\text{BG}}|\mathbf{r} - \mathbf{r'}|}}{4\pi|\mathbf{r}-\mathbf{r'}|}\mathbf{J(\mathbf{r'})}
$. Eq. 3 is called the electric current density volume integral equation, or JVIE\cite{JVIE}, and allows us to determine the scattered electromagnetic fields by solving for $\mathbf{J}_{\text{eq}}(\mathbf{r})$: $
\mathbf{E_{\text{scat}}}(\mathbf{r}) = \frac {1}{j \omega \varepsilon_0}(\mathcal{N}-\mathcal{I})\mathbf{J}_{\text{eq}}(\mathbf{r})$, 
$\mathbf{H_{\text{scat}}}(\mathbf{r}) = \mathcal{K}\mathbf{J}_{\text{eq}}(\mathbf{r})$,
where the $\mathcal{K}$ operator is defined as $
\mathcal{K}(\mathbf{J}) = \nabla \times \int \frac{e^{-jk^{BG}|\mathbf{r} - \mathbf{r'}|}}{4\pi|\mathbf{r}-\mathbf{r'}|}\mathbf{J(\mathbf{r'})}$. To solve Eq. 3, the environment is uniformly discretized into cubic voxels; then, by applying the Galerkin method with spatial piecewise-constant basis functions (note that higher-order basis functions can also be used if desired, e.g., \cite{linear_basis}), Eq. 3 is transformed into a matrix equation\cite{JVIE}:
\begin{equation}
    (I - MN)J = J_{\text{inc}}
\end{equation}
Here, M is a diagonal matrix representing the discretized $\mathcal{M}$ operator and contains the material information for each voxel in the domain, N is a Block Toeplitz with Toeplitz Blocks (BTTB) matrix representing the $\mathcal{N}$ operator, and $J_{\text{inc}} =  j \omega \varepsilon_0M E_{\text{inc}}$ (further details on discretizing the JVIE are provided in the Supporting Information, section S1). Eq. 4 can be solved using an iterative Krylov subspace solver, such as the generalized minimal residual method (GMRES)~\cite{GMRES}. For a system of n voxels, although the system matrix (I - MN) is dense, FFT-based MVP techniques can be used due to the BTTB structure of N to reduce the computational cost of the matrix-vector product (MVP) from $O(n^2)$ to $O(n \log n)$~\cite{JVIE}; Moreover, circulant preconditioners have been shown to be highly effective in reducing the number of solver iterations, without significantly increasing the MVP time\cite{Circ_Prec}. When using an iterative linear system solver, the FFT-accelerated MVP combined with superior conditioning provides the JVIE with a distinct speed advantage over other frequency-domain solvers, such as FEM and FDFD, particularly for complex 3D systems with a large number of voxels.

To utilize the JVIE solver to design practical nanophotonic devices, mode sources are required to launch unidirectionally propagating modes in the waveguide without exciting other undesired modes. This can be achieved by employing appropriate equivalent electric and magnetic surface current densities on a plane to satisfy the electromagnetic transmission boundary conditions (further details are provided in the Supporting Information, section S2). As an example, as shown in Fig. 1a, a straight waveguide with silicon ($Si$) core material and silicon dioxide ($SiO_2$) cladding, with a $500 \, \text{nm} \times 225 \, \text{nm}$ cross-section, is simulated under fundamental mode excitation. To simulate the waveguide extending to infinity, adiabatic absorbers\cite{Absorbers} are used at both ends, which have a similar effect to using a Perfectly Matched Layer (PML) in FD methods. Figs. 1b and 1c show the magnitude and the real part of the y component of the electric field for the fundamental mode, respectively. As can be seen, the electromagnetic wave propagates in only one direction and is absorbed without reflection at the end of the domain, indicating accurate fundamental mode excitation and propagation.
\begin{figure}
    \centering
    \includegraphics[width=1.0\linewidth]{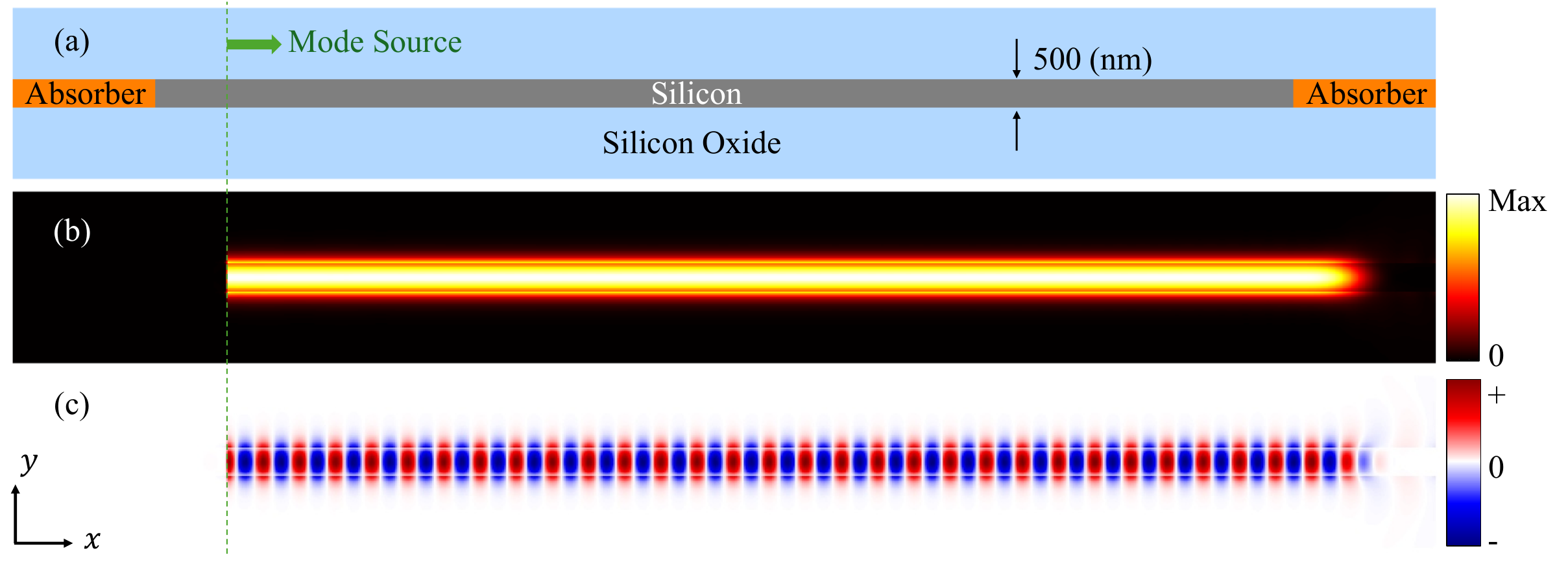}
    \caption{Straight silicon waveguide under fundamental mode excitation. (a) Simulation setup: The mode source is enforced on a plane perpendicular to the propagation direction (shown in green), and the thickness of the structure is 225 nm. (b) Magnitude of the electric field. (c) Real part of the y-component of the electric field. }
    \label{fig:waveguide-fundamental-mode}
\end{figure}

\subsection{Comparison with Finite-Difference Methods}

As mentioned in the previous section, the FFT-based MVP and the use of suitable preconditioners enable JVIE to achieve faster convergence compared to other finite-difference methods, such as FDFD and FDTD, particularly for large 3D structures. To investigate this advantage, a comparison between the JVIE and these two other FD-based methods is presented for the same simple example considered in the previous section of a straight silicon waveguide in a homogeneous silicon dioxide background medium. The open-source SPINS~\cite{SPINS} solver and commercial Lumerical~\cite{Lumerical} solver are used for the FDFD and FDTD simulations, respectively. Different lengths of waveguide are simulated while keeping the constant transverse cross-section, allowing for the study of how the runtime grows as the waveguide length increases. CPU-based iterative solvers are used for both the JVIE and FDFD frequency-domain solvers to solve their corresponding linear systems of matrix equations.

\begin{figure}
    \centering
    \includegraphics[width=1.0\linewidth]{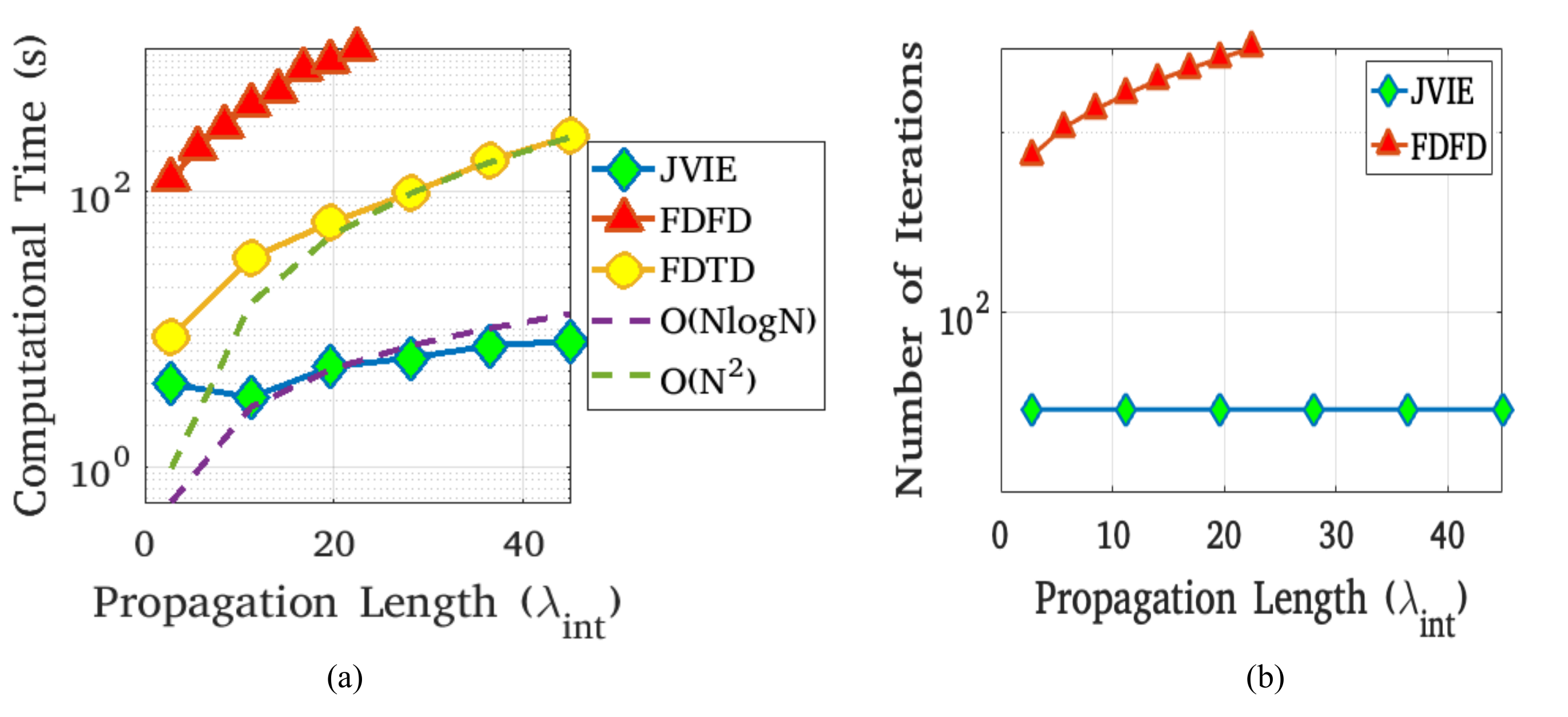}
    \caption{(a) Runtime comparison of the JVIE, commercial FDTD, and FDFD solvers vs. length of the straight silicon waveguide considered in Fig. 1. (b) Number of iterations to achieve GMRES convergence ($10^{-4}$ residual tolerance) for the JVIE and FDFD solvers vs. different waveguide lengths.}
    \label{fig:waveguide-comparison}
\end{figure}

\begin{figure}
    \centering
    \includegraphics[width=1.0\linewidth]{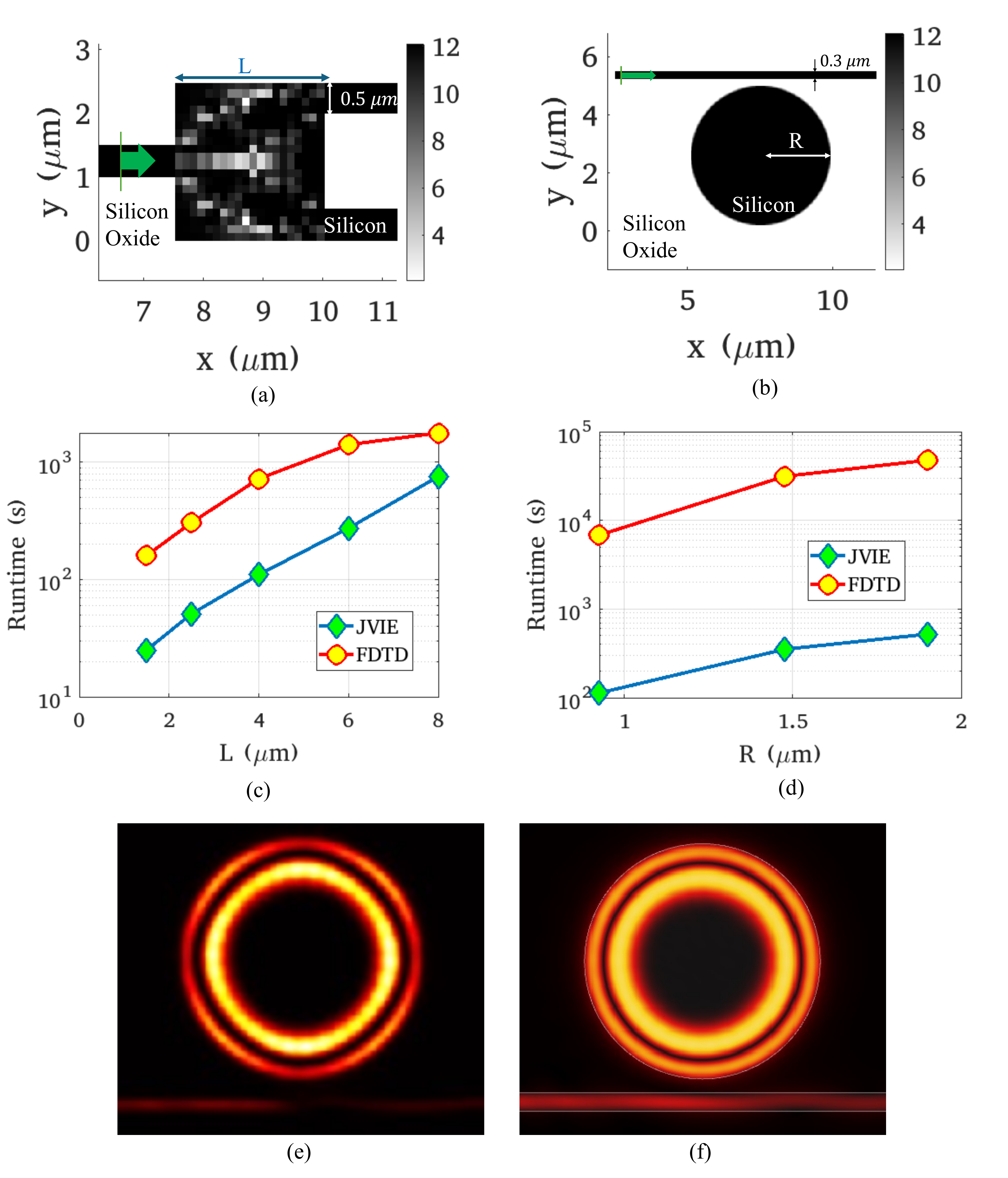}
    \caption{The permittivity of the (a) The power splitter structure and (b) the Microdisk structure. Runtime comparison of the JVIE and FDTD solvers for (c) the power splitter with gray pixels for different lengths of L, and for (d) the microdisk for different values of R. Magnitude of the magnetic field |\textbf{H}| for the resonant mode for $R = 1.9 \mu m$  achieved from the (e) JVIE solver at the wavelength $\lambda = 1.582 \mu m$ and (f) FDTD solver at the wavelength $\lambda = 1.590 \mu m$.}
    \label{fig:resonator-comparison}
\end{figure}

Fig. 2a presents a comparison of the computation times for the three different solvers. JVIE outperforms FDTD and FDFD for all the lengths considered and enjoys the slowest rate of increase in computing time vs. propagation length. For longer waveguides, the runtimes of the finite-difference solvers become substantially (several orders of magnitude) higher than those of the JVIE. Additionally, convergence of the iterative solver could not be achieved for the FDFD solver for any of the cases longer than 25 wavelengths. Reference lines for $O(N \log N)$ and $O(N^2)$ are shown in Fig. 2a, in which $N$ is the number of simulation grid points, to further illustrate the speed advantages of the JVIE method. Fig. 2b demonstrates that the number of iterations required by the JVIE solver remains constant as the length of the waveguide increases, owing to the effective preconditioning approach employed for its corresponding matrix equation; however, the number of iterations required by the FDFD solver starts off several orders of magnitude higher and quickly blows up further as the waveguide grows longer.

To evaluate the performance of JVIE for more complex structures, we compare the solution times for a structure with gray-scale features and a microdisk resonator against the commercial FDTD solver. Fig. 3a illustrates a power splitter structure, comprising an input waveguide that launches the fundamental mode, a central region composed of pixels with continuous permittivities between those of silicon and silicon oxide, which contains a multitude of gray-scale features, and two output waveguides. The input and output waveguides have a $0.5 \mu m \times 0.225 \mu m$ cross-section, and the simulation grid size for both JVIE and FDTD is 25 nm. First-level circulant preconditioners are applied to the waveguides, and a second-level circulant preconditioner is applied to the central region. The central region size is $L\times L$, and the runtime of both solvers is shown in Fig. 3c for different lengths of $L$. As shown in Fig. 3c, although the runtime of JVIE increases as $L$ increases, it remains smaller than that of the FDTD solver for every case considered.

As another example, as shown in Fig. 3b, a microdisk resonator is simulated when its resonant mode is excited at its resonant wavelength by a bus waveguide with a $0.3\mu m\times0.4\mu m$ cross-section. The excited mode is TE, and using the same grid size of 25 nm for both solvers, the runtime of the JVIE and FDTD solvers is measured for different radii of the microdisk, as shown in Fig. 3d. A first-level circulant preconditioner is applied to the bus waveguide, and a second-level circulant preconditioner is applied to the disk in the JVIE solver. To more accurately represent the curvature of the microdisk (as shown in Fig. 3b) on a Cartesian voxel grid, dielectric averaging was used by applying a Gaussian smoothing kernel~\cite{Gaussian_Filter}. The difference in resonant wavelength between our dielectric averaging approach and that used by the FDTD solver was less than 10 nanometers. The structure in each solver was excited at resonance for 4 different radii. Figs. 3e, and 3f show the magnitude of the magnetic field for $R = 1.9 \mu m$ achieved from the JVIE and FDTD solvers, respectively. Their computational times are measured and shown in Fig. 3d. The times required to reach convergence of the FDTD solver are significantly larger than those of the JVIE, as expected, since the FDTD solver must run enough time-steps for the energy to escape the resonator, resulting in a prolonged simulation time and slow convergence. On the other hand, since JVIE is a frequency-domain solver, it does not suffer from this problem. Thus, we expect that it can be very useful for simulating devices with resonances or other energy-trapping structures that cannot be handled easily via other frequency-domain approaches. The timing of the FDFD solver for these large structures was not included due to being unable to reach convergence in a reasonable amount of time.

In conclusion, JVIE proves to be a faster solver compared to finite-difference solvers for large-scale 3D simulations, making it a promising alternative simulation technique for the inverse design of nanophotonic devices.

\section{Inverse Design of Nanophotonic Structures}

Three nanophotonic devices are designed using the JVIE solver to demonstrate its potential for inverse design: a 3dB power splitter, a dual-wavelength Bragg grating, and a selective mode reflector. Note that the comprehensive description of the selective mode reflector example is provided in the Supporting Information, section S4. In general, the optimization process includes defining a cost function that represents the performance of the structure, calculating the gradient of the cost function with respect to the design parameters, and finally, using an appropriate gradient-based optimizer to adjust the design parameters in order to minimize the cost function and improve the device's performance.

For the 3dB power splitter and the selective mode reflector, we employ topology optimization, wherein a specific region is defined as the optimization domain and divided into pixels. The goal is to find an optimized pixel pattern that maximizes the performance of the structure according to the cost function. To ensure the final design is easy to fabricate, the pixel sizes are chosen to be large ($100 nm \times 100 nm$ for the mode reflector and $125 nm \times 125 nm$ for the power splitter) compared to the operational wavelength. Since the structures consist of pixelized patterns with no sub-pixels or smaller features, the minimum feature size and the minimum gap size for the fabrication equal the pixel sizes (100 nm for the selective mode reflector and 125 nm for the power splitter), which satisfy standard silicon-on-insulator (SOI) nanofabrication constraints. \cite{SubWavelength, Levelset} During the topology optimization phase, the permittivity of each pixel is allowed to vary smoothly between that of the silicon dioxide ($n_{SiO_2}=1.44$) cladding and that of silicon ($n_{Si}=3.46$) core. The optimization unknowns are allowed to vary from $0$ to $1$, which are mapped to $n_{SiO_2}$ and $n_{Si}$ respectively. The final goal is to arrive at a binary design consisting of only $Si$ and $SiO_2$ pixels via suitable binarization techniques. For the dual-wavelength Bragg grating, we employ shape optimization, where the length dimensions of the structure are adjusted and optimized to achieve the desired performance.

At each step of the optimization process, the gradient of the cost function with respect to the design parameters must be calculated. This can be done efficiently using the adjoint method, in which the adjoint problem is solved in addition to the original forward problem at each iteration of the optimization to compute the gradient in a single step regardless of the number of optimization variables (further details on the mathematical derivation can be found in the Supporting Information, Section S3).

We used the gradient-based quasi-Newton optimizer, L-BFGS\cite{LBFGS}, to minimize the cost function and improve the device performance. The topology optimization strategy used consists of two phases: In the first "continuous phase", the optimization parameters are allowed to take on any value between 0 and 1. Although this step can achieve high performance, the resulting structure is not fabricable due to its "grayscale" nature, and a binarization technique must therefore be employed. In the second "discretization phase", the final structure of the first phase is used as the initial structure, and a sigmoid filter is applied to each parameter p to push the optimization parameters toward 0 and 1:
\begin{equation}
    \Tilde{p} = \frac{tanh(\beta\eta)+tanh(\beta(p-\eta))}{tanh(\beta\eta)+tanh(\beta(1-\eta))}
\end{equation}
where $\eta = 0.5$. A low initial value (such as 2) is chosen for the slope ($\beta$) to have a smooth transition, and the L-BFGS optimizer is run to convergence. Next, $\beta$ is increased and L-BFGS is rerun with the previous solution as the starting point, and this process is repeated for larger and larger values of $\beta$ until the filter is strong enough ($\beta \approx 30$) to produce a near binary final pattern\cite{Discrete_Step}. Finally, the remaining parameters are hard thresholded by setting them to 0 or 1 based on their proximity to these values to obtain the final structure. For the shape optimization, only one continuous phase is required followed by truncating the final lengths to a 5nm grid by rounding to satisfy typical design rule constraints. The performance of the final structure was also validated by using the commercial full-wave FDTD simulator, Lumerical.

\subsection{Power Splitter}
As the first example, we consider the JVIE-based inverse design of a 1:2 power splitter, one of the most essential components in the silicon photonics industry. Fig. 4a illustrates the power splitter structure, which consists of an input waveguide that launches the incoming wave into the device, a central design region measuring 2.5 $\mu$m $\times$ 2.5 $\mu$m, composed of 20 $\times$ 20 pixels, each 125 nm $\times$ 125 nm in size, and two output waveguides that receive the evenly split power. The input and output waveguides have a width of $0.5 \mu m$, and the thickness of the whole structure is $0.225 \mu m$.

\setcounter{figure}{3}
\begin{figure}
    \centering
    \includegraphics[width=1.0\linewidth]{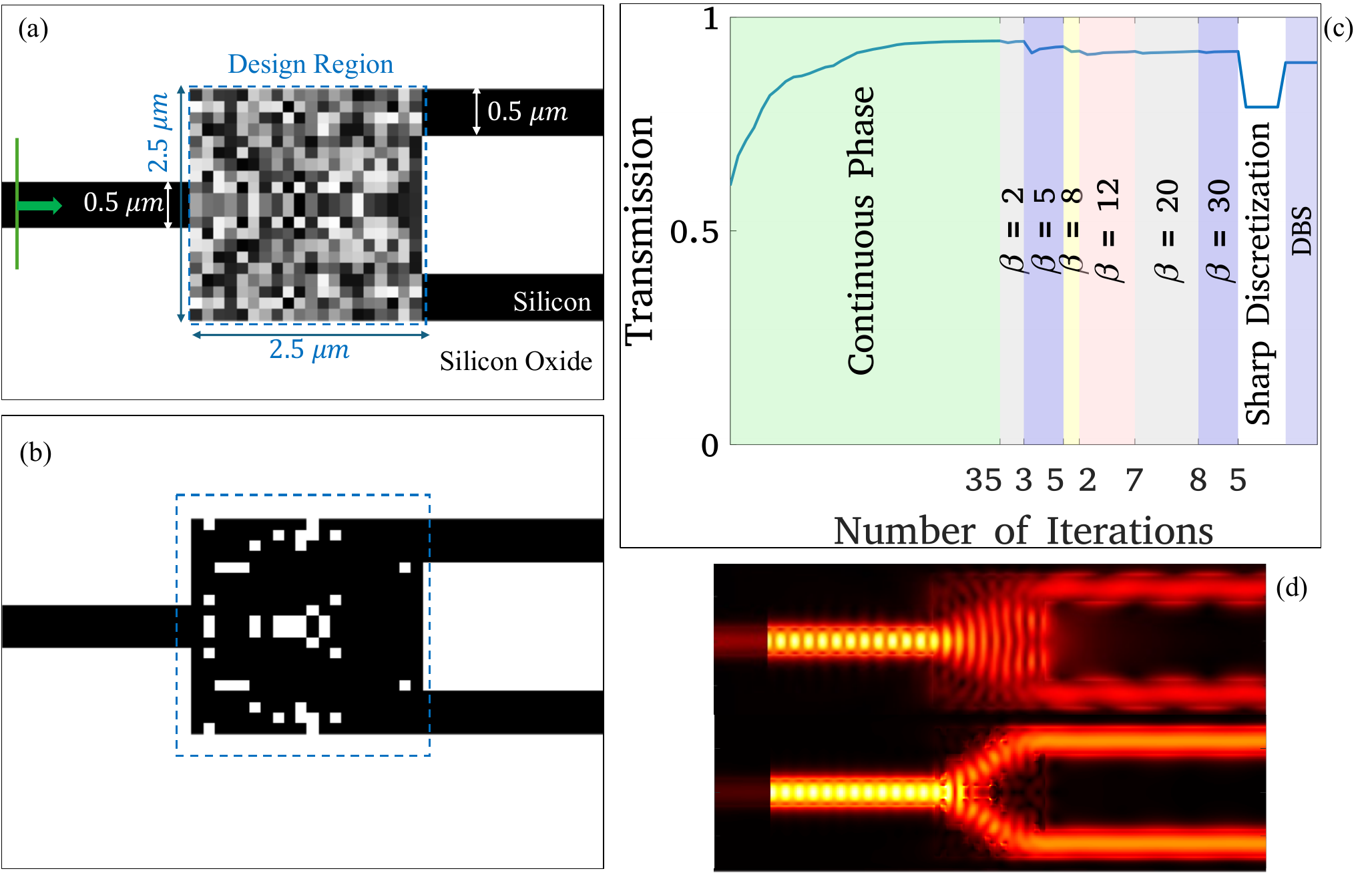}
    \caption{(a) Power splitter simulation setup: The mode source is enforced on a plane perpendicular to the propagation direction (shown in green), and the thickness of the structure is 225 nm. (b) The final optimized structure. (c) Optimization process: The transmission of the fundamental mode at each iteration is shown for both the continuous phase (green shaded area) and the discrete phase (different slopes are shown with different shaded colors). After sharp discretization, Direct Binary Search (DBS) is performed, as shown in the last shaded area, which recovers a significant amount of the performance lost due to thresholding. (d) The magnitude of the electric fields for the all-silicon initial starting point (top) and the final optimized design (bottom). The optimized structure has 0.3 dB insertion loss.}
    \label{fig:power_Splitter_structure}
\end{figure}

The free-space wavelength is 1550 nm, and the side length of each cubic discretization voxel is 25 nm, which is small enough compared to the wavelength inside the core material ($\lambda_{Si}/18$) to ensure an accurate EM solution. As before, circulant preconditioners are applied to the matrix equation, and a fundamental mode excitation source is placed in the input waveguide. A mode monitor is placed on one of the output waveguides to measure the transmission of the fundamental mode. Symmetry is strictly enforced in the design region to split the input power equally between the output ports.

Using $T$ to indicate the transmittance of the fundamental mode in the output waveguide, the cost function can be defined as $f = 0.5 - T$. The L-BFGS optimizer is used with the previously introduced continuous and discrete optimization phases to minimize the cost function and, in turn, increase the transmitted power of the fundamental mode. Fig. 4b shows the final structure, which is a binary, readily fabricable power splitter. Fig. 4c illustrates the optimization process. As with the mode reflector, after the continuous phase, multiple optimizations are done with the sigmoid filter enabled, in which the slope ($\beta$) of the filter is increased after each L-BFGS run until the pattern converges. As can be seen in Fig. 4c, each time the slope increases, the performance degrades, indicating the significant challenges involved with binarizing structures with such large pixels. Ultimately, after sharp thresholding, the final performance of the fully discrete design needed further improvement, and in order to achieve a binary pattern with better performance, we applied the Gradient-oriented Binary Search\cite{Binary_Search} (GBS) algorithm to further optimize the design using discrete optimization.

The structure's performance was validated through simulation using the commercial full-wave software, Lumerical. The insertion loss of the splitter is 0.315 dB, indicating its efficient performance despite the large pixel sizes used. Fig. 4d illustrates the magnitude of the electric field. The upper field distribution corresponds to the initial case, in which all the pixels are silicon, while the lower field distribution represents the final fully optimized design. For the initial case, the efficiency of power transmission from the input port to the output ports is 0.420, while for the final design, it reaches 0.931.

The speed advantage of the JVIE solver is demonstrated by simulating the final structure using both the JVIE and FDTD solvers with identical dimensions (as described in Figure 4a) and grid sizes (25 nm). The runtime of the JVIE solver is 1 minute and 40 seconds, whereas that of the FDTD solver is 7 minutes and 31 seconds, demonstrating the higher speed of the JVIE method.

\subsection{Dual-Wavelength Bragg Grating}
To demonstrate the performance of the JVIE design framework for a larger and more complex structure, a dual-wavelength Bragg grating is designed. Bragg gratings are periodic structures with corrugations (as shown in Fig. 5a) that reflect the components of the incoming waves with a wavelength of $\lambda_B = 2n_{eff}\Lambda$ and transmit the components of other wavelengths. $\Lambda$ is the periodic length, and $n_{eff}$ is the effective refractive index in the grating. In a uniform, conventional Bragg grating, in which all the segments are the same, only one wavelength is reflected in a window of interest. Nevertheless, adding more reflecting channels can improve the Bragg grating's applications in wavelength division multiplexing (WDM) \cite{Bragg_WDM} and optical sensing \cite{Bragg_Sensor}. As a result, a number of prior works have focused on adding more stop bands to the Bragg grating\cite{Bragg_NarrowB, Bragg_Analytical}. In this example, by apodizing the corrugation in each segment, a dual-wavelength Bragg grating is designed, which is a long structure that poses a significant challenge for an FDTD solver to simulate and optimize.

Fig. 5a shows the grating structure, in which the input and output waveguides have a width of $0.5\mu m$, and the thickness of the whole structure is $0.225\mu m$. The grating includes 100 periodic segments, each with a length of $0.65\mu m$, resulting in a total length of $65\mu m$. The material of the structure is silicon, surrounded by silicon oxide as the background material. As shown in the inset of Fig. 5a, each segment consists of a corrugation with a width of $w_i$ and a height of $h_i$. As a result, our goal is to optimize the values of $w_i$ and $h_i$ for $i=1$ to $100$, so that the grating has two reflecting wavelengths.
Unlike our previous designs, we employ shape optimization, in which the parameters $w_i$ and $h_i$ are adjusted relative to the gradient of an objective function with respect to them. Since the design parameters are continuous lengths, as shown in Fig. 5b, some of the simulation grid voxels (sized $dx\times dx$) may contain both silicon and silicon oxide materials, where $w\times h$ is the area of the silicon. Volumetric dielectric averaging is used to calculate the effective permittivity of these voxels as: $\varepsilon = \varepsilon_{Si} \frac{wh}{{dx}^2} + \varepsilon_{SiO_2} (1-\frac{wh}{{dx}^2} )$. A mode source is used to excite the fundamental mode in the input waveguide, and a mode monitor is placed behind it to capture the reflected power and compute the reflection coefficient. The voxel size $dx$ is set to 25 nm, and a first-level circulant preconditioner is applied to accelerate the JVIE solver.

The optimization is performed for $ 1.565 \mu m$ and $ 1.535 \mu m$, both of which are within the optical C-band. Therefore, the objective function is defined as $f = (1 - R_1) + (1 - R_2)$, where $R_1$ and $R_2$ are the reflected power at wavelengths $1.565\mu m$ and $1.535 \mu m$, respectively. In the optimization process, due to optimizing at two wavelengths, two forward simulations and two adjoint simulations are done to compute the gradient with respect to the design parameters. The initial design is a uniform grating whose parameters are chosen to maximize the reflection at $1.565\mu m$, although the reflection is near zero at $1.535\mu m$. Fig. 5c shows the reflections during the optimization process, which converges after 25 iterations. To obtain parameters reasonable for fabrication, the final $w$ and $h$ parameters are truncated to a 5 nm grid resolution~\cite{Bragg_Apd}. The final parameter values are represented in Fig. 5d. Fig. 5e shows the reflection spectrum of the initial structure and the optimized structure, in which the initial spectrum only exhibits a peak at $ 1.565 \mu m$, and the final structure has two peaks at $1.565\mu m, 1.535 \mu m$, corresponding to a dual-wavelength Bragg grating as intended. Figs. 5f and 5g show the magnitude of the magnetic field at $1.565\mu m,1.535 \mu m$ wavelengths, respectively. The top image in each panel corresponds to the initial uniform grating, in which the fields are strongly reflected only at $1.565\mu m$, while there is little reflection at $1.535\mu m$. The bottom images correspond to the optimized grating, in which it can be seen that the fields are strongly reflected at both $1.565\mu m$ and $1.535\mu m$.

To compare the speed advantage of the JVIE solver with the FDTD solver, the final structure is simulated in both solvers with the same dimensions and the same grid resolution (25 nm), and the runtimes are measured. Due to the large size of the grating, the FDTD solver requires significant time to reach convergence since many time steps are required for the fields to propagate to the end of the structure. However, the JVIE solver, due to being a frequency-domain method with desirable conditioning properties, converges much faster. The runtimes of JVIE for simulations at $1.565\mu m,1.535 \mu m$ are 2 minutes and 43 seconds and 2 minutes and 58 seconds, respectively, corresponding to a total runtime of 5 minutes and 41 seconds. Although a single FDTD simulation can be used to obtain the response at both wavelengths, it takes 2 hours and 23 minutes to converge. As a result, the JVIE design framework is 25 times faster for the Bragg grating example than the FDTD solver, highlighting its efficacy for designing large structures that may be intractable using alternative full-wave solution methods.

\begin{figure}
    \centering
    \includegraphics[width=1.0\linewidth]{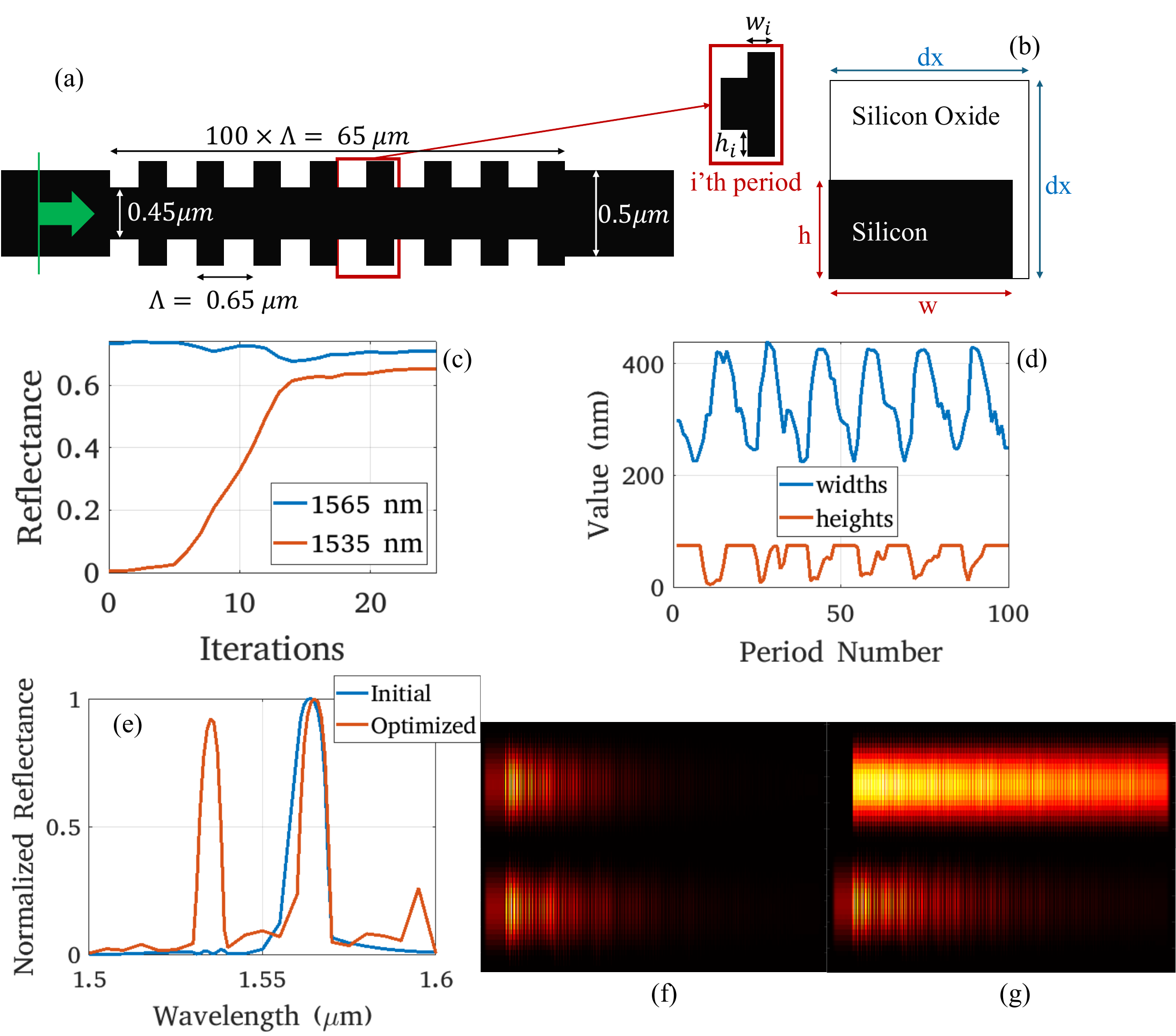}
    \caption{(a) Bragg grating's structure. (b) Continuous material averaging (c) Reflectance at wavelengths $1.565\mu m,1.535 \mu m$ during the optimization process. (d) The optimized parameters' values. (e) The reflectance spectrum of the initial structure and the optimized structure. (f) The magnitude of the magnetic fields for the initial structure (top) and the optimized structure (bottom) at $1.565\mu m$ wavelength. (g) The magnitude of the magnetic fields for the initial structure (top) and the optimized structure (bottom) at $1.535\mu m$ wavelength.}
    \label{fig:placeholder}
\end{figure}

\section{Discussion and Conclusion}
This article explores and demonstrates the viability of the JVIE, a fast, full-wave 3D electromagnetic solution method, for the inverse design of nanophotonic devices. Gradients of the objective function with respect to the design parameters are efficiently computed using the adjoint method and used in a gradient-based optimizer. When comparing against finite-difference-based methods, JVIE shows notable advantages for the inverse design of large-scale, practical structures that are long in one dimension, resulting in significantly reduced simulation times. This indicates the significant benefits of using the JVIE to inverse design such devices. The three silicon photonic design examples considered in this work all utilized large cubic pixels, thereby maximizing robustness and fabricability at the expense of a more challenging inverse design process. The simulated performance of the devices is comparable to that of other state-of-the-art power splitters~\cite{Splitter_1, Splitter_2}, dual-wavelength Bragg gratings~\cite{Bragg_NarrowB, Bragg_MW}, and selective mode reflectors~\cite{Vuckovic_Mode_Ref} reported in the literature.

This paper primarily focused on demonstrating the feasibility and effectiveness of the proposed frequency-domain volume integral equation-based topology and shape optimization platform. For future work, despite adding more unknowns and complexity to the system, incorporating higher-order basis functions may result in faster convergence and higher accuracy with coarser grids, especially for structures with high-index contrasts \cite{linear_basis}. Moreover, the circulant preconditioner becomes less effective for highly inhomogeneous structures, resulting in more GMRES iterations required to achieve convergence and correspondingly slower runtimes, which may limit the design of more complicated structures. As a result, investigating more advanced preconditioners and/or fast direct matrix solution methods could further enhance the JVIE-based design framework. Furthermore, we expect that significant additional speedups can be achieved in future work by extending the JVIE solver to a GPU-based hardware-accelerated implementation. To apply this approach to design larger, more complex devices, curvilinear meshing strategies~\cite{Curvilinear} could be utilized and alternative algorithmic acceleration algorithms~\cite{IFGF} could also be leveraged to further increase the speed of the process. Additionally, in topology optimization, it is possible for the optimizer to converge to a local optimum, especially when the pdesign roblem is large and complex. Therefore, it may be required to optimize many structures with different random initial starting points to find a device with acceptable performance. As a result, especially for complex design problems, the inverse design strategy could be further enhanced by also employing more advanced optimization algorithms, as well as machine learning methods, to better explore the optimization landscape and accelerate convergence. \cite{Hilab}

Finally, it is worth mentioning that the method could also be applied to broader applications, such as plasmonic structures containing metals, in which the permittivity of a metal can be modeled by adding an imaginary term, as well as multilayer geometries, which can be simulated using the JVIE solver by representing the permittivities of each layer with voxelized grids. Furthermore, magnetic materials could also be modeled with a slight extension to the formulation by introducing an additional magnetic current density, albeit at the cost of doubling the number of computational unknowns.

\section*{Associated Content}

\subsection*{Supporting Information}
Supporting Information Available: Details of the JVIE formulation and discretization, as well as methods used in the optimization process, including mode source excitation and derivation of gradients with respect to the design parameters are provided in the Supporting Information. The design process of a selective mode reflector is also presented as an additional example using the JVIE-based optimization framework.

\section*{Funding}
The authors gratefully acknowledge support by the Air Force Office of Scientific Research (FA9550-25-1-0020) and the National Science Foundation (CCF-2047433).

\bibliography{references}


\clearpage

\renewcommand{\thepage}{S\arabic{page}}
\renewcommand{\theequation}{S\arabic{equation}}
\renewcommand{\thefigure}{S\arabic{figure}}
\renewcommand{\bibnumfmt}[1]{(S#1)}
\renewcommand{\citenumfont}[1]{S#1}
\setcounter{page}{1}
\setcounter{equation}{0}
\setcounter{figure}{0}
\begin{center}
  \textbf{\huge Supplementary Information}
\end{center}

\tableofcontents

\newpage

\section{JVIE discretization}
\addcontentsline{toc}{section}{JVIE discretization}

To discretize the JVIE formulation, we employ the Galerkin method using piecewise-constant basis functions. The domain is divided into uniform cubic voxels of size $\Delta\times\Delta\times\Delta$, with $N_x$, $N_y$, and $N_z$ voxels along the x, y, and z axes, respectively. Defining $m_x = 1,2,..,N_x$, $m_y = 1,2,..,N_y$, $m_z = 1,2,..,N_z$, and the 3D voxel index $\mathbf{m} = (m_x,m_y,m_z)$, the basis functions are given by:
\begin{equation}
    f_\mathbf{m}(x,y,z) =
    \begin{cases} 
        1, & \text{if } (m_x - 1)\Delta < x < m_x\Delta, \\
           & \quad (m_y - 1)\Delta < y < m_y\Delta, \\
           & \quad (m_z - 1)\Delta < z < m_z\Delta, \\
        0, & \text{otherwise}.
    \end{cases}
\end{equation}

The equivalent electric volume current density can then be approximated as:
\begin{equation}
    \mathbf{J}_\text{eq}(x,y,z) = \sum_{\mathbf{m}} \alpha^x_{\mathbf{m}}f_\mathbf{m}(x,y,z) \hat{x} + \sum_{\mathbf{m}} \alpha^y_{\mathbf{m}}f_\mathbf{m}(x,y,z)\hat{y} + \sum_{\mathbf{m}} \alpha^z_{\mathbf{m}}f_\mathbf{m}(x,y,z)\hat{z} 
\end{equation}

where $\alpha^x_{\mathbf{m}}, \alpha^y_{\mathbf{m}}$, and $\alpha^z_{\mathbf{m}}$ are the unknown coefficients of the system. Thus, the total number of unknowns is $3\times Nx \times Ny \times Nz$. 

Applying the Galerkin method to test the system by performing the inner product on both sides of the JVIE against the same set of basis functions results in the discrete matrix equation:
\begin{equation}
    (I - MN) J = J_{\text{inc}}
\end{equation}

Here, $I$ is the identity matrix, and $M$ is a diagonal matrix, where each diagonal element is given by:
\begin{equation}
    M_{\mathbf{m},\mathbf{m}} = \frac{\varepsilon_r(\mathbf{r}) - \varepsilon_r^{BG}}{\varepsilon_r(\mathbf{r}) }
\end{equation}

where $\mathbf{m} = (m_x,m_y,m_z)$ represents the 3D voxel index, and its corresponding position vector $\mathbf{r} = (m_x - \frac{1}{2}) \Delta~ \hat{x} + (m_x - \frac{1}{2}) \Delta~ \hat{y} + (m_x - \frac{1}{2}) \Delta~ \hat{z}$ \textsuperscript{S1}. It is worth mentioning that if $N_{voxel}$ denotes the number of voxels, the matrix $M$ has size $3N_{voxel}\times3N_{voxel}$. Although the JVIE can support anisotropic media, since all the structures considered in this work are isotropic, the elements are repeated 3 times on the diagonal of $M$ \textsuperscript{S1}.

The N operator is structured as:
\begin{equation}
    N = 
    \begin{bmatrix}
        N^\text{xx} & N^\text{xy}  & N^\text{xz}  \\
        N^\text{yx} & N^\text{yy}  & N^\text{yz}  \\
        N^\text{zx} & N^\text{zy}  & N^\text{zz} 
    \end{bmatrix}
\end{equation}

Each element of $N$ is given by \textsuperscript{S1}:
\begin{equation}
    N^{\alpha\beta}_{\textbf{m}\textbf{n}} = \sum_k \sum_l (\hat{n}_k \times \hat{\alpha}).(\hat{n'}_l \times \hat{\beta}) \int_{s_k}\int_{s_l} \frac{e^{-jk_0|\textbf{r} - \textbf{r}' - \textbf{r}_m - \textbf{r}'_n|}}{4\pi|\textbf{r} - \textbf{r}' - \textbf{r}_m - \textbf{r}'_n|}dsds'
\end{equation}
where \textbf{m}, \textbf{n} are 3D voxel indices, $\textbf{r}_m$, $\textbf{r}_n$ represent the positions of voxels \textbf{m}, \textbf{n}, respectively, and $s_k$, $s_l$ correspond to one of the six faces of the voxels \textbf{m} and \textbf{n}, respectively. 

By closely examining Eq. S5, it can be observed that each of the nine submatrices $N^{\alpha\beta}$ is a BTTB matrix. Consequently, the matrix-vector product (MVP) of $N$ with a vector x of size n can be efficiently computed using a 3D Fast Fourier Transform (FFT), reducing the computational cost to $O(N \log N$). Since $M$ is a diagonal matrix and costs only $O(N)$ to apply, the MVP of the entire system matrix also has $O(N\log N)$ asymptotic complexity.

\section {Methods}
\addcontentsline{toc}{section}{Methods}

In this section, details of mode source excitation and gradient calculation in the JVIE framework are discussed, which are the key components in the JVIE-based optimization process.

\subsection{Mode source in JVIE solver}

Fig. S1a shows the mode source plane in the waveguide. In the discretized JVIE formula (Eq. 3 of the main text), the incident equivalent current density is expressed as $J_{\text{inc}} = j\omega\varepsilon_0ME_{\text{inc}}$, where $E_{\text{inc}}$ is the vector of the incident electric field in the absence of the dielectric scatterer. Thus, as shown in Fig. S1b, to excite a mode, the source should generate the corresponding electromagnetic fields in the background medium, which, in our case, is silicon dioxide ($SiO_2$).

As shown in Fig. S1b, to ensure that the excited electromagnetic fields propagate in only one direction, the tangential mode fields ($\mathbf{E}_\text{mode}$, $\mathbf{H}_\text{mode}$) are extracted. Following this, the appropriate electric and magnetic surface current densities are imposed on a surface to satisfy the boundary conditions:
\begin{equation}
    \mathbf{J}_s = \hat{n} \times (\mathbf{H}_\text{mode} - \mathbf{0})
\end{equation}
\begin{equation}
    \mathbf{M}_s = -\hat{n} \times (\mathbf{E}_\text{mode} - \mathbf{0})
\end{equation}
\begin{figure}
    \centering
    \includegraphics[width=1.0\linewidth]{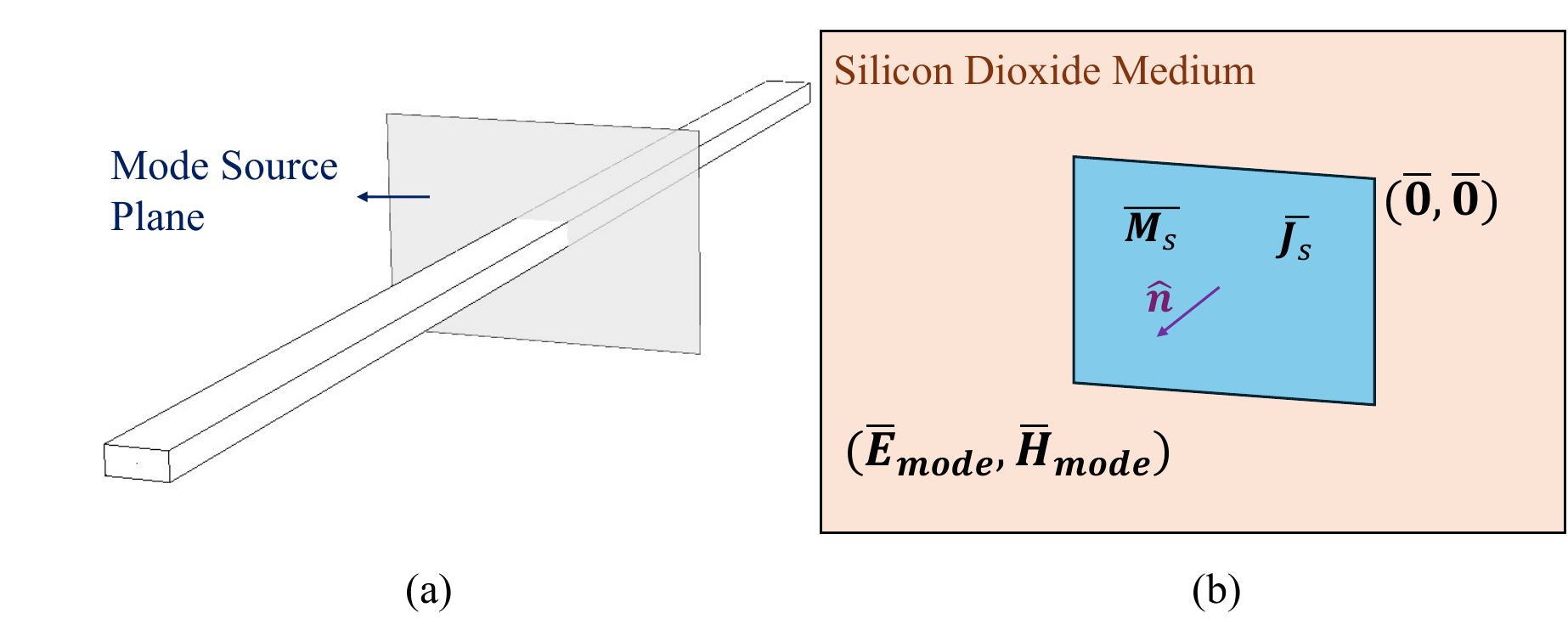}
    \caption{(a) Mode source plane in the waveguide. (b) Generating the excitation electromagnetic fields in the background medium (silicon dioxide) for the JVIE system.}
    \label{fig:mode_source}
\end{figure}

Therefore, using these electric and magnetic surface current densities as sources ensures that the corresponding electromagnetic fields are generated on the right-hand side, while no fields appear on the left-hand side, effectively enforcing unidirectional wave propagation.

To calculate the generated electromagnetic fields by these sources, one can employ the integral equations derived in Ref. S2; however, it is important to consider that the right-hand side of the discretized JVIE equation involves the inner product between the continuous incident electric field $\mathbf{E}_{\text{inc}}$ and the basis functions. This requires additional integral calculations, which are computationally expensive.

A better approach is to treat $\mathbf{J}_s$, $\mathbf{M}_s$ as effective electric and magnetic current densities and use the JVIE operators to compute the generated fields directly:
\begin{equation}
    {E_{\text{inc}}} = \frac {1}{j \omega \varepsilon_0}(N-I)J_s - KM_s
\end{equation}
\begin{equation}
    {H_{\text{inc}}} = \frac {1}{j \omega \mu_0}(N-I)M_s + KJ_s
\end{equation}

This makes the approximation that the effective source densities $\mathbf{J}_s$ and $\mathbf{M}_s$ are constant over each discrete 3D voxel on the simulation grid. Since the same $N$ and $K$ operators are used from the $JVIE$ formulation, however, this enables rapid computation of $E_{inc}$ and $H_{inc}$ by leveraging FFTs. As shown in Fig. 1 in the main text, especially for small $\Delta$, not only is this approach significantly faster, but it also maintains sufficient accuracy for practical applications.

\subsection{Gradient calculation using the adjoint method}

In this section, mathematical details for calculating the gradient of the cost function with respect to the design parameters are discussed. 

For the power splitter design, the cost function is defined as $f = 0.5 - T$; where $T = |a_1|^2$ is the transmission of the fundamental mode, measured at a monitor placed perpendicularly in one of the output waveguides, and $a_1$ is the forward modal coefficient of the fundamental mode.

To compute the derivative of the cost function $f$ with respect to the design parameters $p$, the chain rule is applied. Since $f$ is real-valued, while $a_1$ and $J_{\text{eq}}$ are complex, the Wirtinger derivative \textsuperscript{S3} is used:
\begin{equation}
    \frac{df}{dp} = 2Real(\frac{df}{da_1}\frac{da_1}{d_{J_{\text{eq}}}}\frac{d{J_{\text{eq}}}}{d{\varepsilon}}\frac{d{\varepsilon}}{dp})
\end{equation}

where $J_{\text{eq}}$ is the solution of the JVIE solver, and $\varepsilon$ is a vector containing the permittivities of all voxels in the JVIE domain. As a result, each term in Eq. S11 must be mathematically derived to compute the gradient accurately.

Since $f = 1 - |a_1|^2$, its derivative with respect to $a_1$ is given by $\frac{df}{da_1} = -{a_1}^*$.

The modal coefficient $a_1$ is obtained using the mode overlap integral:
\begin{equation}
    a_1 = \frac{1}{4}(\frac{\int{(\mathbf{E} \times \mathbf{H}^*_1).\mathbf{ds}}}{N_1} + \frac{\int{(\mathbf{E}^*_1 \times \mathbf{H}).\mathbf{ds}}}{N_1^*})
\end{equation}
where $\mathbf{E}^*_1$ and $\mathbf{H}^*_1$ are the electric and magnetic fields of the fundamental mode, respectively, while $\mathbf{E}$ and $\mathbf{H}$ are the electric and magnetic fields at the monitor, computed by the JVIE solver. The normalization factor $N_1 = \frac{1}{2}\int{(\mathbf{E}_1 \times \mathbf{H}^*_1).\mathbf{ds}} $ is the complex power of the fundamental mode.

The electric and magnetic fields obtained from the JVIE solver consist of both the incident and scattered components. Consequently, $a_1$ can be expressed as:
\begin{equation}
    a_1 = \frac{1}{4}(\frac{\int{(\mathbf{E}_s \times \mathbf{H}^*_1).\mathbf{ds}}}{N_1} + \frac{\int{(\mathbf{E}^*_1 \times \mathbf{H}_s).\mathbf{ds}}}{N_1^*}) +  \frac{1}{4}(\frac{\int{(\mathbf{E}_{\text{inc}} \times \mathbf{H}^*_1).\mathbf{ds}}}{N_1} + \frac{\int{(\mathbf{E}^*_1 \times \mathbf{H}_{\text{inc}}).\mathbf{ds}}}{N_1^*})
\end{equation}

in which $a_{\text{inc}} = \frac{1}{4}(\frac{\int{(\mathbf{E}_{\text{inc}} \times \mathbf{H}^*_1).\mathbf{ds}}}{N_1} + \frac{\int{(\mathbf{E}^*_1 \times \mathbf{H}_{\text{inc}}).\mathbf{ds}}}{N_1^*})$ is a constant term. By applying the trapezoidal quadrature rule to the integrals, the discretized form of the remaining term can be expressed as ${c_E}^TEs + {c_H}^THs$, where $Es$ and $Hs$ are the discrete scattered field vectors on the solution grid, and $c_E$ and $c_H$ are coefficient vectors that have absorbed both the quadrature weights and constant modal fields for the $E_s$ and $H_s$ integrals, respectively

Moreover, $E_s = \frac {1}{j \omega \varepsilon_0}(N-I)J_{\text{eq}}$ and
$H_s = KJ_{\text{eq}}$. Therefore, $a_1$ can be expressed as:
\begin{equation}
    a_1 = c^TJ_{\text{eq}} + a_{\text{inc}}
\end{equation}
where $c = \frac {1}{j \omega \varepsilon_0}c_E(N-I) + c_HK$ is a constant vector. As a result, the forward modal coefficient can be directly obtained from the solution of the JVIE solver by simply computing a vector-vector dot product. Moreover, the derivative of $a_1$ with respect to $J_{\text{eq}}$ is given by $\frac{da_1}{dJ_{\text{eq}}} = c^T$.

Next, we consider the derivative $\frac{d{J_{\text{eq}}}}{d{\varepsilon}}$. Eq. S15 represents the JVIE formulation, explicitly showing the dependence of the matrices and vectors on $\varepsilon$:
\begin{equation}
    (I - M(\varepsilon)N)J(\varepsilon) = j \omega \varepsilon_0M(\varepsilon)E_{\text{inc}}
\end{equation}

By differentiating Eq. S15 with respect to $\varepsilon_i$, the relative permittivity of voxel $i$, we obtain:
\begin{equation}
    -\frac{dM(\varepsilon)}{d\varepsilon_i}NJ(\varepsilon) + [1 - M(\varepsilon)N]\frac{dJ(\varepsilon)}{d\varepsilon_i} = j \omega \varepsilon_0\frac{dM(\varepsilon)}{d\varepsilon_i}E_{\text{inc}}
\end{equation}

and thus,

\begin{equation}
    \frac{dJ(\varepsilon)}{d\varepsilon_i} = {[1 - M(\varepsilon)N]}^{-1}\frac{dM(\varepsilon)}{d\varepsilon_i}[j \omega \varepsilon_0E_{\text{inc}} + NJ(\varepsilon)]
\end{equation}

Defining $d_i = \frac{dM(\varepsilon)}{d\varepsilon_i}[j \omega \varepsilon_0E_{\text{inc}} + NJ(\varepsilon)]$, we obtain:

\begin{equation}
    \frac{dJ(\varepsilon)}{d\varepsilon} = {[1 - M(\varepsilon)N]}^{-1}[d_1 d_2 ... d_{N_{voxel}}]
\end{equation}

where $N_{voxel}$ is the number of voxels in the JVIE domain.

Finally, $\frac{d{\varepsilon}}{dp}$ is computed based on how the design parameters are mapped to the permittivity of the voxels. Assuming that the operator Q performs the spatial mapping of the design pixels onto the JVIE voxels, the permittivity distribution is given by: 

\begin{equation}
    \varepsilon = \varepsilon_{\text{BG}} + (\varepsilon_{\text{core}} - \varepsilon_{\text{BG}}) Qp.
\end{equation}

Therefore, differentiating with respect to $p$, we obtain $\frac{d{\varepsilon}}{dp} = (\varepsilon_{\text{core}} - \varepsilon_{\text{background}}) Q$.

In conclusion, the gradient of the cost function $f$ with respect to the design parameters $p$ is given by:
\begin{equation}
\begin{split}
    \frac{df}{dp} &= 2Real(\frac{df}{da_1}\frac{da_1}{d_{J_{\text{eq}}}}\frac{d{J_{\text{eq}}}}{d{\varepsilon}}\frac{d{\varepsilon}}{dp})\\ &= 2Real({a_1}^*c^T{[I - MN]}^{-1}[d_1 d_2 ... d_{N_{voxel}}](\varepsilon_{\text{core}} - \varepsilon_{\text{background}}) Q)
\end{split}
\end{equation}

The matrix $(I - MN)$ is the JVIE system matrix, which means that ${[I - MN]}^{-1}d_i$ corresponds to solving the JVIE system with $d_i$ as the excitation. Therefore, computing the gradient that involves evaluating the term ${[I - MN]}^{-1}[d_1 d_2 ... d_N]$ requires solving the JVIE for $N_{voxel}$ times; which is extremely inefficient, especially for large $N_{voxel}$. Nevertheless, the adjoint method enables efficient gradient computation as formulated in Eq. S20, reducing the computational cost to just a single JVIE solve.

By defining ${J_{\text{adj}}}^T = {a_1}^*c^T{[I - MN]}^{-1}$, the gradient in Eq. S20 can be efficiently computed using simple matrix-vector products. The main challenge is computing the adjoint current density vector $J_{\text{adj}}$. Since ${J_{\text{adj}}}^T = {a_1}^*c^T{[I - MN]}^{-1}$, thus, $J_{\text{adj}} = {[I - MN]}^{-T}{a_1}^*c = {[I - NM]}^{-1}{a_1}^*c$ since M is diagonal and the N matrix is self-adjoint owing to using the Galerkin method, which leverages the same basis and testing functions. Therefore, $J_{\text{adj}}$ is obtained by solving a system with the matrix $(I - NM)$ and excitation ${a_1}^*c$, which has the same level of complexity as the original JVIE solution.

The only remaining parts of Eq. S20 are the $d_i$ columns. Based on the definition of the M matrix, the derivative $\frac{dM(\varepsilon)}{d\varepsilon_i}$ is a diagonal matrix with all diagonal elements equal to zero, except for the entries at indices $i$, $N_{voxel} + i$, and $2N_{voxel} + i$, which are given by $\frac{1}{{\varepsilon_i}^2}$. Therefore, we obtain:
\begin{equation}
    d_i = \frac{dM(\varepsilon)}{d\varepsilon_i}[j \omega \varepsilon_0E_{\text{inc}} + NJ] = \frac{1}{{\varepsilon_i}^2}
    \begin{bmatrix}
    0 \\
    \vdots \\
    {[j \omega \varepsilon_0E_{\text{inc}} + NJ]}_i \\
    0\\
    \vdots \\
    0 \\
    [j \omega \varepsilon_0E_{\text{inc}} + NJ]_{i+N} \\
    0 \\
    \vdots \\
    0 \\
    [j \omega \varepsilon_0E_{\text{inc}} + NJ]_{i+2N} \\
    0 \\
    \end{bmatrix}
\end{equation}

As a result, ${J_{\text{adj}}}^T d_i$ is obtained simply as:

    \begin{equation}
    \begin{split}
        {J_{\text{adj}}}^T d_i &= \frac{1}{{\varepsilon_i}^2} \\ 
        &\Big({[j \omega \varepsilon_0E_{\text{inc}} + NJ]}_i {[J_{\text{adj}}]}_i 
        + {[j \omega \varepsilon_0E_{\text{inc}} + NJ]}_{N_{voxel}+i} {[J_{\text{adj}}]}_{N_{voxel}+i} \\
        &+ {[j \omega \varepsilon_0E_{\text{inc}} + NJ]}_{2N_{voxel}+i} {[J_{\text{adj}}]}_{2N_{voxel}+i}\Big)
    \end{split}
\end{equation}

Therefore, the computation of ${J_{\text{adj}}}^T [d_1 d_2 ... d_N]$ requires only $O(N)$ operations. As a result, the gradient can be efficiently calculated using Eq. S20.

Although the cost function for the selective mode reflector design, $f = R_2 - R_1$, is different, the gradient calculation process follows a similar process. Here, $R_1 = {|b_1|}^2$ and $R_2 = {|b_2|}^2$ are the reflected powers of the fundamental mode and the second mode, respectively, where $b_1$ and $b_2$ are the backward modal coefficients of these modes. These coefficients are obtained using the mode overlap integrals:

\begin{equation}
    b_1 = \frac{1}{4}(\frac{\int{(\mathbf{E} \times \mathbf{H}^*_1).\mathbf{ds}}}{N_1} - \frac{\int{(\mathbf{E}^*_1 \times \mathbf{H}).\mathbf{ds}}}{N_1^*})
\end{equation}
\begin{equation}
    b_2 = \frac{1}{4}(\frac{\int{(\mathbf{E} \times \mathbf{H}^*_2).\mathbf{ds}}}{N_2} - \frac{\int{(\mathbf{E}^*_2 \times \mathbf{H}).\mathbf{ds}}}{N_2^*})
\end{equation}

where $\mathbf{E}^*_1$ and $\mathbf{H}^*_1$ are the electric and magnetic fields of the fundamental mode, and $\mathbf{E}^*_2$ and $\mathbf{H}^*_2$ are the electric and magnetic fields of the second mode. The fields $\mathbf{E}$ and $\mathbf{H}$ are obtained from the JVIE solver at the monitor. The normalization factors, $N_1 = \frac{1}{2}\int{(\mathbf{E}_1 \times \mathbf{H}^*_1).\mathbf{ds}} $ and $N_2 = \frac{1}{2}\int{(\mathbf{E}_2 \times \mathbf{H}^*_2).\mathbf{ds}} $ represents the complex powers of the fundamental and second modes, respectively.

Similar to $a_1$, the backward modal coefficients $b_1$ and $b_2$ can be expressed as ${c_1}^TJ_{\text{eq}} + b^1_{\text{inc}}$ and ${c_2}^TJ_{\text{eq}} + b^2_{\text{inc}}$, respectively. Following a similar derivation, the gradients of $R_1 = {|b_1|}^2$ and $R_2 = {|b_2|}^2$ with respect to the design parameters are given by:

\begin{equation}
    \frac{dR_1}{dp} = 2Real({b_1}^*{c_1}^T{[I - MN]}^{-1}[d_1 d_2 ... d_{N_{voxel}}](\varepsilon_{\text{core}} - \varepsilon_{\text{background}}) Q)
\end{equation}

\begin{equation}
    \frac{dR_2}{dp} = 2Real({b_2}^*{c_2}^T{[I - MN]}^{-1}[d_1 d_2 ... d_{N_{voxel}}](\varepsilon_{\text{core}} - \varepsilon_{\text{background}}) Q)
\end{equation}

\begin{equation}
    \frac{df}{dp} = \frac{dR_2}{dp} - \frac{dR_1}{dp}
\end{equation}

Again, to compute the gradients in Equations S25 and S26 efficiently, the adjoint electric current densities can be defined as ${J_{\text{adj},1}}^T = {b_1}^*{c_1}^T{[I - MN]}^{-1}$ and ${J_{\text{adj},2}}^T = {b_2}^*{c_2}^T{[I - MN]}^{-1}$, respectively, and obtained by running the corresponding solver for the adjoint system.

\section{Selective Mode Reflector Inverse Design}
\addcontentsline{toc}{section}{Selective Mode Reflector Inverse Design}

As an example of the JVIE design framework, a selective mode reflector is designed to reflect a desired mode while scattering out all undesired modes. This structure is useful in nanophotonic devices that are prone to generating undesired, higher-order modes, and their performance is dependent on forcing these modes out of the system. For instance, as shown in Fig. S2a, to achieve a high quality factor (Q), a Fabry-Perot resonator's waveguide is designed to be wider than usual, which could result in the generation of higher-order modes. Placing such selective mode reflectors at both ends scatters these modes ($TE_{10}$ and other unwanted modes) out of the resonator while acting as mirrors for the desired, fundamental mode ($TE_{00}$); thus, only the fundamental mode is enhanced inside the resonator.\cite{ResonatorC}

\begin{figure}
    \centering
    \includegraphics[width=1.0\linewidth]{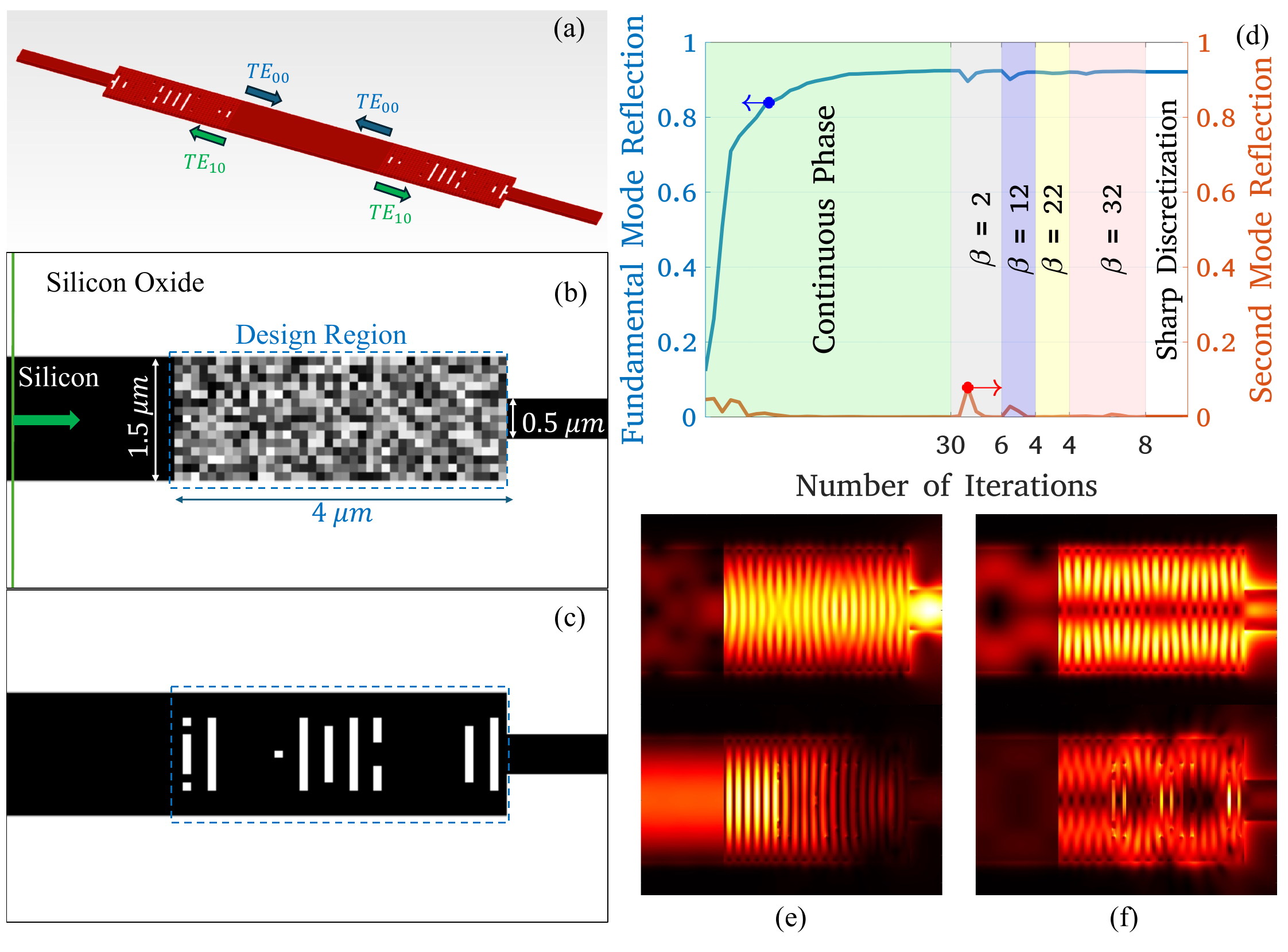}
    \caption{(a) Fabry-Parot resonator with selective mode reflectors on each end. (b) Selective mode reflector simulation setup: The mode source is enforced on a plane perpendicular to the propagation direction (shown in green), and the thickness of the structure is 225 nm. (c) The optimized pattern of the mode reflector. (d) Optimization process: Transmissions of both modes in each iteration are shown for both the continuous phase (green shaded area) and the discrete phase (different slopes are shown with different shaded colors). (e),(f) Magnitude of the electric field under the excitation of the fundamental mode and the second mode, respectively. The top row of plots corresponds to the simple all-silicon initial design, and the bottom row corresponds to the final optimized pattern. The transmissions of the fundamental mode and the second mode for the optimized design are 0.953 and 0.003, respectively.}
    \label{fig:mode_reflector_final}
\end{figure}
Fig. S2b illustrates the selective mode reflector structure. Silicon and silicon dioxide are used as the core and cladding materials, respectively. The waveguide on the left side resembles the wide waveguide in the resonator, with a width of 1.5$\mu$m, and the waveguide on the right side is narrower with a width of 0.5$\mu$m. The optimization design region in the center has a size of 4 $\mu$m $\times$ 1.5 $\mu$m and is discretized into a grid of 40 by 15 pixels of size 100 nm $\times$ 100 nm. The design objective is to find an optimal binary pattern for the pixels in the design region such that the fundamental mode is completely reflected while the other modes are transmitted. The thickness of the whole structure is $0.225 \mu m$

The free-space wavelength is 1550 nm, and each cubic voxel's length in the VIE discretization is 25 nm per side, which is small enough compared to the wavelength inside the core material ($\lambda_{Si}/18$) to ensure an accurate field solution. Circulant preconditioners are applied to the matrix equation, and two mode sources, the fundamental mode ($TE_{00}$) source and the second mode ($TE_{10}$) source, are inserted on the left waveguide. Moreover, a mode monitor is placed behind the mode sources to measure the reflected power of each mode.

The cost function is defined as $f = R_2 - R_1$, where $R_1$ and $R_2$ are the reflectances of the fundamental mode and the second mode, respectively. The L-BFGS optimizer is used in both the continuous and discrete phases to minimize the cost function, and therefore, increase the reflectance of the fundamental mode while reducing the reflectance of the second mode. Fig. S2c shows the final structure, which is a binary, readily fabricable design. Furthermore, the optimization process is illustrated in Fig. S2d, including both the continuous and the discrete phases. After the continuous phase, the sigmoid filter is enabled, and for each slope ($\beta$), we rerun the optimizer using the final design of the previous run as the initial starting point of the next one. As we increase the slope, the performance recovers fast, as seen in Fig. S2d; thus, the discrete phase is a smooth process. In the end, by applying a sharp filter, we achieve a fully binary structure with acceptable performance.

To validate its performance, the structure is also simulated using the commercial full-wave simulator Lumerical. Figs. S2e and S2f show the magnitude of the electric field under fundamental mode and second mode excitation, respectively. The upper field distributions correspond to the initial solution, where all pixels are silicon, while the field plots directly below represent the optimized pattern. It is evident that, unlike the initial case, the optimized pattern reflects the majority of the fundamental mode power while fully transmitting the second mode power. The reflectances of the fundamental mode and second mode are 0.953 and 0.003, respectively, demonstrating excellent performance despite the use of relatively large pixel sizes.

To demonstrate the speed advantage of the JVIE solver, we simulate the optimized structure using both the JVIE and FDTD solvers with the same dimensions (as described in Figure 3b) and grid sizes (25 nm), and measure their respective runtimes. The JVIE solver’s runtimes for the fundamental mode and the second mode simulations are 1 minute and 37 seconds and 1 minute and 34 seconds, respectively. In comparison, the runtimes for the FDTD solver are 12 minutes and 2 seconds and 10 minutes and 16 seconds. Therefore, the JVIE solver has a significant time-saving advantage over the FDTD solver, especially when used for inverse design, in which the forward and adjoint solvers are called numerous times before achieving convergence.

\end{document}